\documentclass[12pt]{iopart}
\usepackage{graphicx}
\usepackage{setstack}
\usepackage{iopams}
\usepackage{color}

\def\th{\theta}
\newcommand{\be}{\begin{equation}}
\newcommand{\ee}{\end{equation}}
\newcommand{\bea}{\begin{eqnarray}}
\newcommand{\eea}{\end{eqnarray}}

\begin{document}

\title{Optical conductivity of curved graphene}

\author{A J Chaves$^1$, T Frederico$^1$, O Oliveira$^{1,2}$, W de Paula$^{1,3}$ and M C Santos$^2$}
\address{$^1$ Dep. de F\'\i sica, Instituto Tecnol\'ogico de Aeron\'autica, DCTA, 12228-900 S\~ao Jos\'e dos Campos, Brazil}
\address{$^2$ Dep. de F\'\i sica, Universidade de Coimbra, 3004-516 Coimbra, Portugal}
\address{$^3$ Dep. de F\'\i sica, Universidade Federal de S\~ao Carlos, 13565-905 S\~ao Carlos, Brazil}
\ead{tobias@ita.br}

\begin{abstract}
We compute the optical conductivity for an out-of-plane deformation in graphene using an approach based on solutions of the Dirac equation in curved space. Different examples of periodic deformations along one direction translates into an enhancement of the optical
conductivity peaks in the region of the far and mid infrared frequencies for periodicities $\sim100\,$nm.
The width and position of the peaks can be changed by dialling the parameters of the deformation profiles.
The enhancement of the optical conductivity is due to intraband transitions and the translational invariance breaking
in the geometrically deformed background. Furthemore, we derive an analytical solution of the Dirac equation in a curved space for a general deformation along one spatial direction. For this class of geometries, it is shown that curvature induces an extra phase in the electron wave function, which can also be explored to produce interference devices of the Aharonov-Bohm type.
%
%
\end{abstract}


\maketitle

\section{Introduction and Motivation}

The dynamics of electrons in graphene can be described by tight-binding models or, equivalently, by
a two-dimensional Dirac equation\cite{Neto2009,Vozmediano2010,Peres2010,Chaves2011,Oliveira2011,Popovici2012,Cordeiro2013}.
Graphene is a two dimensional material embedded in three dimensions and is assumed to define a 2D flat surface.
However, from the general description of stability analysis for 2D materials and membranes, see e.g.
\cite{Vozmediano2010} and references there in, one expects deviations from flatland. Ripples, with a size distribution of 50 -- 100 \AA,
were observed in suspended graphene \cite{Meyer2007}.
Furthermore, atomistic simulations using many-body interatomic potentials \cite{Fasolino2007} show that ripples
appear spontaneously owing to thermal fluctuations. The predicted size distributions of the ripples were
in good agreement with the experimental observations as shown in \cite{Meyer2007}. Also, structural corrugation
have been observed in graphene \cite{Ishigami2007,Stolyarova2007}.

The deviations of the bulk graphene from a flat surface can also be used to model topological defects
\cite{Katanaev1992,Juan2007}. In this approach, curvature and torsion are associated with disclinations and
dislocations of the medium (see e.g. \cite{Vozmediano2010}).
The connection between the geometrical approach, elasticity theory and tight-binding models was
investigated, for example, in \cite{Juan2012}. A geometric framework considering curvature \cite{Gonzalez1993}
has also been applied to model the electronic properties of the ``buckyball" C$_{60}$. This work provides an example of how
a geometrical approach can be used to address electronic properties.

Curvature in a geometrical approach can be translated into a pseudo magnetic field or, more generally, into
pseudo gauge fields which can lead to observable phenomena; see e.g the references
\cite{Vozmediano2010,GuineaNature2010,Levy2010,Juan2011}.

In \cite{GuineaNature2010} the authors suggested an experiment to measure zero-field quantum Hall
effect in strained graphene. The strain being at the origin of a strong pseudo magnetic field. Furthermore,
they are able to connect strain with the possibility of opening energy gaps in the graphene electronic
spectrum.

Experimental evidence of strain-induced large pseudo magnetic fields were reported in \cite{Levy2010},
where the Landau levels were measured by scanning tunelling microscopy. The authors
claim the presence of pseudo magnetic field in graphene with intensities greater than 300 tesla.

Another example of possible effects due to the pseudo magnetic fields associated with local deformations of
the graphene is given in \cite{Juan2011}, where the authors propose a device to measure
Aharonov-Bohm interferences at the nanometre scale. The interplay between the pseudo-magnetic field  and an external real one
has been addressed in \cite{FariaPRB2013}, where it was shown that due to an inherent competition
between the two fields, the electric current exhibited spatial inhomogeneity not present when the real magnetic field is zero. For flat graphene,
a nonlinear magnetisation appears in the absence of gap or when it is much smaller than the separation between two Landau levels, as has been
shown in \cite{SlizPRB2012}, which one could in principle combine with geometric deformations and study the competition between the pseudo-
and real magnetic fields in the magnetisation.

The fabrication of wrinkled graphene sheets and the measurement of the single layer topological features
by atomic force microscopy was achieved in \cite{Scniepp2006}.
Graphene sheets can be reversibly folded and unfolded multiple times, as reported in a following work \cite{Scniepp2008}.
Although reference \cite{Scniepp2008} was focused in the structural properties related to defects in graphene,
the electronic and transport properties of the folded graphene do not reproduce those of the ``flat" graphene. A periodic deformation in suspended graphene was achieved in \cite{Bao2009}, where the amplitude and the periodicity of the deformation can be controlled by the boundary conditions and temperature. Another approach to engineer periodic deformations in graphene was studied in \cite{Wang2011}, where graphene was deposited on stretchable elastomeric substrates.
In \cite{Xu2009} the electrical properties of graphene were investigated by scanning tunnelling microscopy. They
have identified wrinkles in graphene sheet, with the local curvature of the wrinkle being responsible for the breaking of the lattice
symmetry, and showed that the wrinkles have a lower electrical conductance and have mid gap
states. It is appropriate to recall that structural deformations can induce zero-field quantum Hall effects in strained graphene
\cite{Vozmediano2010}.

In the literature one can find several theoretical works, see e.g.
\cite{Vozmediano2008,Guinea2010,Pereira2010,Pellegrino2010,Zhang2011,Kerner2012},
where the geometrical deformations of the graphene sheet change the electronic and optical
properties of the pristine material.
The optical conductivity for strained graphene (in-plane displacements) was calculated in
\cite{Pereira2010,Pellegrino2010} using a tight-binding approach. In the present work, we rely on
a quantum field theoretical approach with curvature to compute the effect of out-of-plane deformations on the
graphene optical conductivity. To the best knowledge of the authors, this is the first time where the connection
between out-of-plane deformation and graphene optical conductivity is investigated and computed.

In the recent experiment in~\cite{Koch2012} a single graphene nanoribbon was bended and its
voltage-dependent conductance measured. Further, the experiment accounted for the conductance dependence
on the precise atomic structure and bending of the molecule in the junction. This type of bending is a geometric deformation in
one direction.

All the reported works rise the question of how curvature can be used to tailor the electronic and optical properties of graphene
and nanoribbons.
In this sense, it is interesting to explore different geometries of the graphene sheet. A possible application that can explore
the geometric deformations of graphene is, for example, flexible electronics.
The main goal of the current work, is
the theoretical investigation of one dimensional deformations and how they can change the electronic and optical
properties of graphene materials. Experimental examples of such type of deformations
are given by ripples \cite{Meyer2007} or bending \cite{Koch2012}.

Our starting point is the Dirac equation in curved space in 2D+1 dimensions. For a given geometry, we study how stretching
and out-of-plane displacements can be included in a geometrical framework. We provide general expressions that can be
used to investigate any geometry.
Furthermore, we investigate how the electronic dispersion relation, the electronic wave function
and optical conductivity depend on the geometry of the graphene. Several one dimensional out-of-plane periodic deformations are
worked in detail.  We also calculated the expression of the operator velocity for an eletron in a static metric.

We found an analytical solution of the Dirac equation in a general curved surface for deformations along a single space direction.
The effects of the curvature appear as an extra phase in the electron wave function, besides its normalization, but keeps
the usual linear dispersion relation $E = \pm |\vec{k}|$.
This result adds to the work of \cite{Juan2011} the possibility of engineering Aharonov-Bohm interference type devices with
one dimensional geometric deformations of the graphene sheet. For the particular case of periodic deformations
the phase structure implies the quantisation of the electron energy, mimicking Bloch waves in a crystal.

The extra phase and the normalisation due to the geometry of the graphene sheet, change the optical conductivity in a non-trivial way.
Our analytical solutions simplifies considerable the computation of the optical conductivity via the Kubo formula adapted to the
two dimensional case. It also
makes easier the optimisation of the geometry to tune the optical properties.

For one dimensional
out-of-plane deformations, the Dirac equation can be mapped into a Sturm-Liouville type of equation, i.e. a Schr\"odinger
like equation with a non-trivial potential bearing a complex functional form. In this way, the solution of the Sch\"odinger
eigenvalue equation with a non-trivial potential is mapped into a solvable geometrical problem. We detailed such mapping for
a particular out-of-plane geometric deformation of the graphene sheet.

The paper is organised as follows. In section \ref{curved_space}, we review the properties of the Dirac formalism in curved space.
The hamiltonian of the system and the pseudo-gauge fields are defined.

In \ref{sub:displacements1} we discuss how strain and out-of-plane displacements can be accommodated in a geometrical
framework. In \ref{Sec:InPlane}, the solution of the Dirac equation for in-plane displacements are discussed.
In \ref{Sub:1Dgeneral} the analytical solution of the Dirac equation for out-of-plane displacements is derived
for a general one dimensional deformation. Moreover, the geometric phase is defined and its relation to a generalised
Aharonov-Bohm interference device is presented for an idealised situation.
In \ref{Sub:ripples}, the one dimensional periodic ripple is investigated in detail and the quantisation condition is resolved.
The Sturm-Liouville type of equation is derived and the corresponding small and large deformations limits studied.
In section \ref{opt_cond}, the optical conductivity is computed for our analytical solution of the Dirac equation via the Kubo formula.
The large frequency limit will be discussed and, for this limit, we derive analytically forms for the optical conductivity
which take into account the geometric of the graphene sheet. In addition, the optical conductivity is calculated for three geometric
out-of-plane deformations.
Finally, in section \ref{ultima_sec} we resume and conclude. In \ref{Ap:out}  expressions to handle the
general case of out-of-plane deformations are provided.

\section{Dirac Electron in Curved Space \label{curved_space}}

The action for a relativistic electron in a 2D+1 curved space is given by
\begin{equation}
 S = \int d^3x ~ \sqrt{g} ~ \overline\Psi (x) \Big\{
 i \Gamma^{\mu} D_{\mu} - M \Big\} \Psi (x) \, ,
\end{equation}
where $\Psi (x)$ is the Dirac field and $g = \mathrm{det}(g_{\mu\nu})$, with $g_{\mu\nu}$ being the metric tensor.
The Dirac matrices in curved space-time read
\begin{equation}
   \Gamma^{\mu}=e^{\mu}_{A}\gamma^{A} \ ,
\end{equation}
where the ``vielbein" $e^{A}_	{\mu}$ define a local Lorentzian frame such that the space-time interval can
be written as
\begin{equation}
 ds^2 = \, \eta_{AB} \,  \theta^{A} \, \theta^{B} \ ,
\end{equation}
with $\eta_{AB} = \mathrm{diag}( 1, -1, -1)$ being the Minkowski metric and $\theta^{A} = e^A_\mu \, dx^\mu$. The covariant derivative
is
\be
   D_{\mu} = \partial_{\mu} + \frac{1}{4}\omega^{AB}_{\mu}\sigma_{AB}
\ee
with
\be
  \sigma_{AB} = \frac{1}{2} \Big(\gamma_{A}\gamma_{B}-\gamma_{B}\gamma_{A}\Big) \ ,
\ee
and the spin connection
\bea
\omega_{\mu}^{AB} &=& \frac{1}{2} \, e^{\nu A} \, \Big(\partial_{\mu}e^{B}_{\nu} \, - \, \partial_{\nu}e^{B}_{\mu}\Big)\nonumber\\
 &&     \, - \, \frac{1}{2}e^{\nu B}\Big(\partial_{\mu}e^{A}_{\nu} \, - \, \partial_{\nu}e^{A}_{\mu}\Big)\nonumber\\
 &&    \, - \,  \frac{1}{2}e^{\rho A}e^{\sigma B}\Big(\partial_{\rho}e_{\sigma C}
      \, - \, \partial_{\sigma}e_{\rho C} \Big) e^{C}_{\mu}.
\eea
The Dirac equation in a curved space-time is
\be
  \Big\{ i \, \Gamma^{\mu} D_{\mu} - M \Big\}\Psi=0
  \label{Eq:Dirac}
\ee
and the Dirac Hamiltonian for a time independent metric reads
\be
  H = \int d^2x ~ \sqrt{g} ~ \overline\Psi \Big \{ - i \Gamma^j D_j + M \Big\}\Psi \ .
  \label{Eq:Hdef}
\ee

In this particular case one can choose
$\theta^0  = dt$ and $\Gamma^0 = \gamma^0$, where $\gamma^0$ is the usual flatland gamma matrix. Furthermore, from the definition
\bea
 \theta^1 & = &  e^1_1 \, dx ~ + ~ e^1_2 \, dy \ , \nonumber\\
 \theta^2 & = &  e^2_1 \, dx ~ + ~ e^2_2 \, dy \ ,\nonumber\\
\eea
and taking the symmetric solution $e^1_2  = e^2_1$, it follows, after some algebra,
that the only non-vanishing spin connections are $\omega^{12}_i = - \omega^{21}_i$ for $i = 1$, $2$.
%

The Hamiltonian (\ref{Eq:Hdef}) can then be rewritten as
\be
  H = H_1 + H_2 + H_3,     \label{Eq:H}
\ee
where

\bea
H_{1} &=& \int d^2x ~ \Psi^\dagger \Big \{ - i \sqrt{g} \sum^2_{i,j = 1} \alpha_i \, e^j_i \, \nabla_j \Big\}\Psi \label{H1} \ ,
         \\
H_{2} &=& \int d^2x ~ \Psi^\dagger \Big \{ - i \sqrt{g} \sum^2_{i=1} \alpha_i A_i \Big\}\Psi \label{H2} , \\
H_{3} &=& \int d^2x ~ \Psi^\dagger \Big \{ \sqrt{g} M \beta \Big\}\Psi  \ ,\label{H3}
\eea
the Dirac matrices are
\be
 \beta = \gamma^0 \ , \qquad \alpha_i = \beta \gamma^i \ ,
\ee
and the pseudo-gauge fields $A_i$ are defined, in terms of the vielben and spin connections,
as
\be
  A_i  =     \sum^2_{j,k=1} \, \frac{ \epsilon_{ij} }{ 2 } \, e^k_j \, \omega^{12}_k \ ,
  \label{Eq:pseudo_gauge}
\ee
where $\epsilon_{ij}$ is the two dimensional Levi-Civita symbol.

The physical interpretation of the $H_1$, $H_2$ and $H_3$ can be found in
\cite{Juan2007,Neto2009,Juan2012,Manes2013}.
In particular, the liaison between the continuum approach and a generalized tight-binding Hamiltonian
without breaking the symmetries of graphene cristal structure can be found in
\cite{Manes2013}.
The term $H_1$, which reproduces the usual Dirac Hamiltonian for a flat geometric, gives rise to a
position-dependent Fermi velocity. Possible experimental implications rooted in $H_1$ can be found
in e.g. \cite{Juan2007,Juan2012}.
The term $H_2$ looks like a geometric gauge field, up to an $i$ factor, when compared to the usual
definition coming from the minimal U(1) coupling. It induces a Dirac cone shift in momentum space -- see
\cite{Manes2013}. The term $H_3$ is a mass term multiplied by the determinant of the metric
tensor. In the language of field theory, $M \sqrt{g}$ can be viewed as a scalar field.

The properties of graphene can be traced to the hybridization of one $s$ orbital and two in-plane $p$ orbitals leading to
the formation of $\sigma$ bands, and the remaining $p_z$ orbital builds $\pi$ bands. The latter determines the
electronic properties of flat graphene. A deformation of the graphene sheet change the overlap of the various orbitals
and, therefore,
change the electronic dynamics. Strain, or in-plane deformations, keep the relative orientation of the various
orbitals and its main effect is essentially a rescale of the various parameters characterizing the intensity of the
interactions. On the other hand,
out-of-plane deformations change the relative orientation of the $p_z$ orbital, modifying the overlap
between the various atomic orbitals, and, in principle, can lead to new types of interaction.
For example, as summarized in \cite{Neto2009} out-of-plane deformations can lead to the introduction
(in the continuum field theoretical approach) of a new vector-like field.
These new interactions, connected with the out-of-plane deformation,
are associated with second order derivatives of the deformation (see e.g. \cite{Neto2009,Manes2013}),
while the in-plane deformation are associated with first order derivatives. Therefore,
one expects that  these new interactions, coming from bending $p_z$ orbitals, give subleading contributions
to the properties of graphene. In the present work we will consider only the interactions coming from the
geometric deformation of the flat Hamiltonian and will ignore any possible new interaction.


In the current work, we focus on time independent metrics and, for this class of metrics, it can be shown that the
Hamiltonian (\ref{Eq:H}) is Hermitian \cite{Huang2009}. In this way, one ensures that the $H$ eigenvalues are real numbers.

\subsection{Strain and out-of-plane displacements \label{sub:displacements1}}

Let us consider that a flat graphene surface lies on the $xOy$ plane. If graphene is not flat, it defines
the surface
\begin{equation}
   z = h(x,y) \ .
   \label{Eq:outplane}
\end{equation}
The function $h(x,y)$ parametrizes the out-of-plane displacements and induces curvature.

Strain in graphene can be viewed as a local deformation which can be parametrized by the change of
variables
\begin{equation}
  x \rightarrow x^\prime = \mathcal{X}(x,y)  \qquad\mbox{and}\qquad y \rightarrow y^\prime = \mathcal{Y}(x,y) \ ,
  \label{Eq:strain}
\end{equation}
where in terms of $(x^\prime, y^\prime)$ the metric is euclidean. The transformation defined by
(\ref{Eq:strain}) does not change the surface curvature, in the sense that the 2D Ricci scalar defined on the
graphene sheet is invariant under the change of variables.

The spatial distance between two neighbooring points is given by
\begin{equation}
  ds^2 = \left( dx^\prime \right)^2 + \left( dy^\prime \right)^2  + dz^2 \ .
\end{equation}
The corresponding metric tensor reads
\begin{equation}
  g_{ij}  = \mathcal{X}_i \mathcal{X}_j \ + \ \mathcal{Y}_i \mathcal{Y}_j \ + \  h_i h_j
\end{equation}
where $\mathcal{X}_i$ means partial derivative with respect to $x_i$.

Let us consider deformations described by
\begin{eqnarray}
 x^\prime & = & x + u_x(x,y) \ , \\
 y^\prime & = & y + u_y(x,y) \ , \\
 z            & = & h(x,y) \ ,
\end{eqnarray}
where $u_x$, $u_y$ and $h$ are the displacement fields.  The corresponding metric tensor is given by
\begin{equation}
  g_{ij}   =  \delta_{ij}  + 2 \, u_{ij}
  \label{Eq:metric}
\end{equation}
where the strain tensor reads
\begin{equation}
   u_{ij} = \frac{1}{2}\left( \partial_i u_j + \partial_j u_i + \sum^2_{k=1} \partial_i u_k  \  \partial_j u_k
    + \left( \partial_i h \right) \left( \partial_j h \right) \right)  .
    \label{Eq:strain+out}
\end{equation}
In the limit of the small deformations and to first order in the in-plane displacements and to second
order in the out-of-plane displacement, the strain tensor  $u_{ij}$ reproduces exactly the strain
tensor considered in \cite{Juan2012}. The difference for a finite deformation is the second order term
in the in-plane displacements $u_i$.

\section{In-plane displacements \label{Sec:InPlane}}

Let us consider the case of in-plane deformations where $z = h(x,y) = 0$. Recall that, in this case
the transformation (\ref{Eq:strain}) does not introduces curvature on graphene and, therefore, the solution
of the Dirac equation can be found in the $(x^\prime, y^\prime)$  reference frame where the electron behaves
as a free particle. The solution in the $(x, y)$ reference is obtained from the one in  $(x^\prime, y^\prime)$ via
a generalized boost as described in \cite{Birrell1982}.

For in-plane displacements one can write the general solution for the Dirac equation (\ref{Eq:Dirac}) in terms of
the vector $\vec{k} = k_x \hat{e}_x + k_y \hat{e}_y$ as
\be
\Psi =
 \Big\{ i \, \Gamma^{\mu} D_{\mu} + M \Big\} \, \chi (t) \, e^{ \pm \,  i \, \vec{k} \cdot \big( \vec{x}  + \vec{u}(x,y) \big) }
 \ee
where
\be
   \chi (t)  = \chi \, e^{\pm i E t}
\ee
and
\be
   E = \sqrt{ |\vec{k}|^2 + M^2 }.
\ee
In two dimensional representation, $\chi$ is a Pauli spinor. We call the reader attention, that our solution is defined
up to a normalization factor.

\section{Out-of-plane displacements \label{Sub:1Dgeneral}}

The Dirac equation for a general out-of-plane deformation of the graphene sheet described by equation
$z = h(x,y)$ is computed in \ref{Ap:out} and reads
\be
\left\{  i \gamma^{0}\partial_{0} + i \Gamma^{k}\left(\partial_k+\frac{1}{2}\omega_{k}^{12}\gamma_1\gamma_2\right) - M\right\}\Psi=0 \, ,
  \label{Eq:DiracEqCurvedGeral}
\ee
where
\be
\Gamma^i = \frac{1}{( \nabla h )^2} \left\{ \epsilon^{ik} h_k \left( \epsilon^{mn}\gamma^m h_n \right)
 + \frac{h^i ~ \left( \gamma^k h_k \right) }{\sqrt{1 + ( \nabla h )^2}} \right\}
\ee
where $h_n = \partial h/ \partial x^n$, $\nabla h = h_x \hat{e}_x + h_y \hat{e}_y$ and $\hat{e}_i$ the unit vector along direction $i$.

The nonvanishing spinconnection components are
\be
 \omega^{ij}_k = - \omega^{ji}_k = \left(- h_i h_{jk} + h_j h_{ki} \right) \tilde\Omega (x,y)
\ee
where
\be
 \tilde\Omega (x,y) =
 \frac{- 1 + \sqrt{1 + h_x^2 + h_y^2} }{\left(h_x^2 + h_y^2\right)\sqrt{1 + h_x^2 + h_y^2}} \ .
\ee

The Dirac equation (\ref{Eq:DiracEqCurvedGeral}) can be rewritten introducing the pseudo gauge field $A_i$ defined in
(\ref{Eq:pseudo_gauge}). However, to keep track of the out-of-plane deformation, parameterised in terms of
the function $h(x,y)$, and see how the deformation impacts on the electronic and optical properties of graphene we will not use
such type of formalism.

In this section we are interested in a general deformation along one direction where $z = h(x)$.
For the particular case under discussion, the Dirac equation simplifies into
\begin{equation}
\left\{ i \gamma^0 \partial_0 + i \frac{\gamma^x \partial_x}{\sqrt{1 + h^2_x}} + i \gamma^y\partial_y - M \right\} \Psi = 0 \, .
\end{equation}
In the Pauli spinor
\begin{equation}
    \Psi =e^{-iE t} \, \left( \begin{array}{c}\varphi_A \\ \varphi_B \end{array}\right) \, ,
\end{equation}
the upper component $\varphi_A$ is associated with the electron localised in one of the graphene sublattices,
which we call sublattice A, while $\varphi_B$ is associated with the other graphene sublattice B.
For the two dimension representation
\be
\gamma^0 =  \sigma_3 , \quad \gamma^1 =  - i \sigma_1 ,\quad\mbox{and}\quad \gamma^2 =  i \sigma_2,
\ee
where $\sigma_k$ are the Pauli matrices, the Dirac equation reads
\bea
 \left( E - M \right) \varphi_A ~ + ~ \frac{\partial_x \varphi_B}{\sqrt{ 1 + h^2_x }} ~ + ~ i \, \partial_y \varphi_B & =  & 0\, ,
     \label{Eq:Dirac_upper} \\
 \left( E + M \right) \varphi_B ~ - ~ \frac{\partial_x \varphi_A}{\sqrt{ 1 + h^2_x }} ~ + ~ i \, \partial_y \varphi_A & =  & 0 .
     \label{Eq:Dirac_down}
\eea
The solution of these equations is given by
\be
\Psi = N_{k,\lambda}  e^{ i k_x \, \int^x dx^\prime \sqrt{ 1 + h^2_x(x^\prime )}} \, e^{i k_y y}\,
     \left( \begin{array}{c} 1 \\ \frac{i\, k_x + k_y}{\lambda E_k + M} \end{array} \right) \, ,
     \label{Sol:geral}
\ee
where the normalisation constant is
\be
N_{k,\lambda}=\sqrt{\frac{E_k+\lambda M}{2 E_k}},
\ee
and
\be
  E_k = \sqrt{|\vec{k}|^2 + M^2} \, , \label{energy}
\ee
$\lambda=+1$ ($-1$) label positive (negative) energy and $\vec{k} = k_x \hat{e}_x  + k_y \hat{e}_y$. The  wave function is determined
up to a global normalisation factor.

The phase in $\Psi$ associated with the curvature can be removed from the
spinor introducing the new gauge field $\vec{C}$, defined on the the graphene sheet, with components
\be
  C_x = k_x \, \sqrt{ 1 + h^2_x} \qquad\mbox{ and }\qquad C_y = 0 \, .
\ee
This gauge field does not produce any magnetic or electric field and, in this sense, can be considered a ``pure-gauge" configuration.
We would like to call the reader attention that $\vec{C}$ is not the pseudo-gauge field $\vec{A}$ defined in Eq. (\ref{Eq:pseudo_gauge})
but can be introduced to remove the new electron wave function phase. For example, while $\vec{A}$ can give rise to
pseudo-magnetic fields \cite{Vozmediano2008,Guinea2010,Zhang2011}, the pure-gauge field $\vec{C}$ does not.

\subsection{Aharonov-Bohm phases }

\begin{figure}[t] 
   \centering
   \includegraphics[scale=0.35]{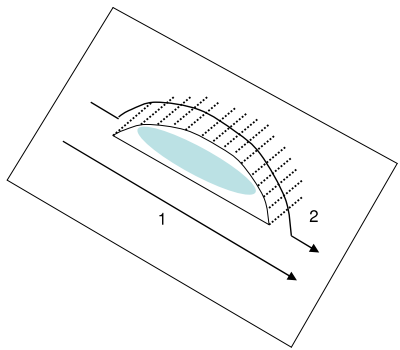}
   \caption{Schematic Aharonov-Bohm interference device.}
   \label{Fig:AharonovBohm}
\end{figure}

The effect of the out-of-plane deformation along one direction translates into a new phase,
\be
    \delta (x) = k_x \int^xdx^\prime \sqrt{ 1 + h^2_x(x^\prime )}  \, ,
    \label{Eq:fase_geometrica}
\ee
in the electron wave function, which can be explored to produce interference devices based on Aharonov-Bohm type effects.
The new phase $\delta (x)$ is a measure of the local curvature.
If in the electron path there are regions of different out-of-plane deformations, being either different lengths or
different metrics, they mimick a double slit Young experiment and the difference of phase between the two path is realised in practice.

In Fig. \ref{Fig:AharonovBohm} we illustrate an idealised Aharonov-Bohm one-dimensional interference device of
a double slit Young type experiment.
Classically, the electron can follow any of the path 1 and 2. If, on the left hand side, the electron is described by the wave
function $\Psi$, in the right side of the apparatus, the electron is described by a superposition of the wave functions, say $\Psi_1$
associated with path 1, and $\Psi_2$ associated with path 2. $\Psi_2$ differs from $\Psi_1$ by a phase $\delta$ given by
Eq. (\ref{Eq:fase_geometrica}). By a convenient choice of the curvature or the length of path 2 one can tune $\delta$ in order to produce
interference effects. A conventional Aharanov-Bohm phase, due to the presence of a solenoid type electromagnetic field,
can be added to the geometric phase $\delta$ and explored to build up or destroy the presence of curvature.

An example of a device relying on the Aharonov-Bohm effect, in association with a rotational symmetric deformation of the geometry,
was proposed in \cite{Juan2011}.
Recently, in \cite{Koch2012} the authors were able to bend a nanoribbon, which could be associated with a metric of the type
discussed here, i.e. $z = h(x)$.
Their technique envisages that, in the future, it may be possible to build an electron multipath arrangement, using
coherently several of their experimental apparatus.

\section{One Dimensional Periodic Out-Of-Plane Deformations  \label{Sub:ripples}}

\begin{figure}[t] 
   \centering
   \includegraphics[scale=0.6]{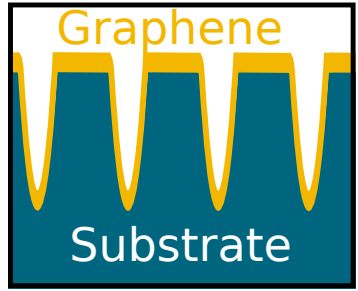}
   \caption{{\it Color on-line.} 1D periodic deformation.}
   \label{perdef}
\end{figure}

  Let us now consider the out-of-plane displacements corresponding to
a periodically corrugated graphene parametrized as (see Fig. \ref{perdef})
\be
  z = h(x)=  h\left(x+\frac{2\pi}{\omega}\right) \, .
  \label{Eq:ripple}
\ee
This type of geometry simulates ripples along the $x$ direction and defines a 1D crystal.
For periodic solutions which require $\delta ( 2 \pi ) = \delta (0) + 2 r \pi$, where $r$ is an integer number,
$k_x$ and, therefore, the energy are quantised. The quantisation condition gives
\begin{equation}
     \frac{k_x}{\omega} \, \xi   =  2\pi \, r\, .
     \label{Eq:quantizacao}
\end{equation}
where $\xi$ is the arc length in units of $1/ \omega$ and is given by
\begin{equation}
     \xi =  \int^{2\pi}_0 d v ~ \sqrt{ 1 + h^2_x ( v) } \ ,
     \label{Eq:quantizacao1}
\end{equation}
computed with the dimensionless argument $v=\omega x$.

\begin{figure}[t] 
   \centering
   \includegraphics[scale=0.35]{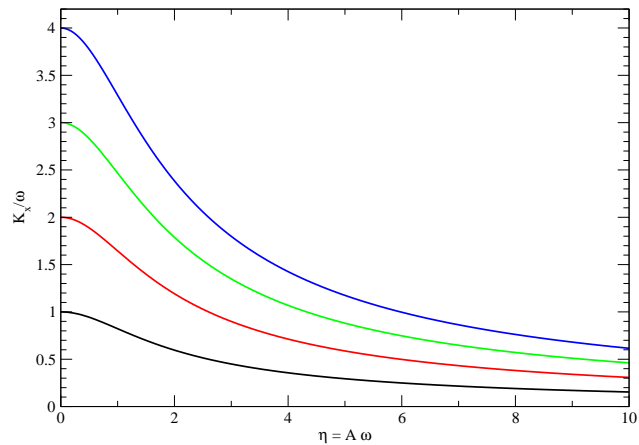}
   \caption{$k_x / \omega$ computed using the quantisation condition (\ref{Eq:quantizacao}) for several
                $r$.}
   \label{Fig:quantizacao}
\end{figure}

For example, for $h(x)= A \cos ( \omega x )$ the quantisation condition (\ref{Eq:quantizacao1}) gives the spectra, as a function
of $\eta = A \omega$, shown in Fig. \ref{Fig:quantizacao}. For small deformations, i.e. for $\eta \ll 1$, Eq. (\ref{Eq:quantizacao}) reads
\begin{equation}
     \frac{k_x}{\omega} = r \left( 1 - \frac{\eta^2}{4}  + \cdots \right) \, .
    \label{Eq:quantizacao_small}
\end{equation}
The quadratic behavior in $\eta$ is clearly visible in Fig. \ref{Fig:quantizacao}.

\subsection{On the Sturm-Liouville equation}

For the geometry given by $h(x)= A \cos ( \omega x )$, the Dirac equation (\ref{Eq:Dirac_upper})--(\ref{Eq:Dirac_down})
can be reduced to a second order differential equation in $\varphi_A$ in the usual way. Indeed,
for a wave function of the form
\be
\varphi_{A}(x,y)=e^{i k_{y}y} \Big\{1+A^2\omega^2\sin^2(\omega x)\Big\}^{1/4}\phi,
\label{phi}
\ee
the second order differential equation in $\varphi_{A}$ can be cast in a Sturm-Liouville form
\be
\left\{ -\frac{d^2 }{dv^2}  + V(v) \right\} \phi (v) = \zeta \, \epsilon(v) \,  \phi (v) \ ,
\label{Eq:SL}
\ee
where $v = \omega x$,
$\zeta= (E^2  - M^2 - k_{y}^2)/\omega^2\ ,$
\bea
\epsilon(v)  &=&  1+\eta^2\sin^2(v)  \qquad\mbox{and}\quad \nonumber\\
&&V(v)   =  - \frac{\eta^2}{16} \,
\frac{ 8 \cos(2v) - \eta^2 \left[8 \sin^2 (v) + \sin^2 (2 v) \right]}
                    {\left( 1 + \eta^2 \sin^2 (v) \right)^2}  \, .
\eea
with $\eta = A \, \omega$. In physical units, the potential energy associated with Sturm-Liouville equation is
given by $\omega^2 \, V(v)$.  To illustrate the Sturm-Liouville form of Dirac equation, we
discuss in \ref{STLE-Mathieu}, the small and large curvature limits, where the eigenvalue equation (\ref{Eq:SL}) reduces to the well-known Mathieu differential equation.

The potential energy which appears in the second order differential equation (\ref{Eq:SL}) is a non-trivial periodic potential and, in
principle, equation (\ref{Eq:SL}) can be solved numerically. However, as shown previously, by looking at the problem from a pure
geometric point of view, one is able to compute the spectrum and the corresponding eigenfunctions easily.
This suggests that one could map complex non-trivial potential energies into geometric problems and look for the solutions of the
corresponding Dirac equation in curved space.  Of course, the question of mapping non-trivial potentials to geometry is, \textit{per see}, a
non-trivial problem.

\section{Optical Conductivity \label{opt_cond}}

In our description of geometrically deformed (out-of-plane) graphene, the active electrons belong to two bands distinguished by the sign of the
corresponding single particle energy eigenstates of the hamiltonian (\ref{Eq:H});
c.f. (\ref{Eq:H_Dirac_geral}) in the first quantised form.
The electron valence levels are the negative energy eigenstates of $H$,
while the conduction band is associated with the positive energy eigenstates of the hamiltonian.

The chemical potential $\mu$ outlines the occupation of the electronic bands. A negative chemical potential
means a partially occupied valence band. On the other hand, a $\mu>0$  describes a partially filled conduction band.

In our  particle independent model, the optical conductivity of deformed graphene can be computed via the Kubo
formula~\cite{Mahan} adapted to a two dimension problem as
\be
\sigma_{ij} = - i \, \mathcal{G} \, \sum_{m,n}\frac{f(E_m-\mu) - f(E_n-\mu)}{\Omega-\Omega_{mn}+i\epsilon} \, v^i_{mn} \, v^j_{nm}, \label{kubo}
\ee
where
\be
 \mathcal{G} = \frac{e^2g_s}{S \Omega} \ ,
\ee
$e$ is the electron electric charge, $g_s = 4$ is the spin and pseudospin degenerescence,
$S$ is the area of the graphene sheet, $\Omega$ is the light frequency, $\Omega_{mn} = E_m - E_n$ is the transition energy,
$f(\varepsilon)=(1+e^{\beta \, \varepsilon})^{-1}$ is the Fermi-Dirac distribution function, $\beta=1/k_BT$ and
$v^i_{mn}=\langle m|v_i| n\rangle$ are the matrix elements of the velocity operator in the Hamiltonian eigenvector basis (\ref{Sol:geral}).

The  velocity operators are defined as $v_i = i \, \left[ H , x_i \right]$, where $H$ is the hamiltonian.
For a general surface geometry $H$ is given in (\ref{Eq:H_Dirac_geral})
and it follows that
\bea
v_i &=& \alpha_k e_{k}^{i} =
   \frac{1}{h_i^2 + h_j^2}
   \left\{ \left( h_j^2 + \frac{h_i^2}{\sqrt{1 + h_i^2 + h_j^2}}\right) \alpha_i \right. \nonumber\\
&& \left. - h_i h_j \left(1 - \frac{1}{\sqrt{1+ h_i^2 + h_j^2}}\right) \alpha_j\right\},  \label{Eq:vel}
\eea
where $\vec\alpha=\beta \vec \gamma$,  $h_{1}\equiv h_{x}$ and $h_{2}\equiv h_{y}$.
The first term in (\ref{Eq:vel}) reproduces the usual operator form for Dirac electrons in flat geometries, i.e. $v_i = \alpha_i$,
up to a geometrical factor which includes the surface deformation, through derivatives of the profile function $h(x,y)$.
The second term add the contribution of a transverse velocity
operator to the chosen direction and vanishes in the limit where $h_i \rightarrow 0$, i.e. a flat surface.

The real part of the longitudinal optical conductivity $\sigma_{ii}$ is associated with absorption and it reads
\bea
\Re \, \sigma _{ii}  &=& - \mathcal{G} \, \pi \, \sum_{m,n}\Big[f(E_m-\mu)-f(E_n-\mu)\Big]\nonumber\\
&&\times\delta\left(\Omega-\Omega_{mn}\right)\,|v^i_{mn}| ^2 \label{kubo1}
\eea
where the sum over states is constrained by energy conservation.
 It can be shown that the  transverse part $\sigma_{i j}$ with $i \ne j$
is proportional to the mass term $M$ or the graphene half-gap. Typical values for $M$ are of the order of meV or less and, therefore,
one can safely disregard the transverse component of the optical conductivity. For example, for a gapless flat graphene
the transverse conductivity vanishes, unless the magnetic field is active \cite{GusJPCM06}.

The electronic transitions associated to optical absorption can occur between levels with different energy signs, named interband
transitions and represented by $\sigma^{\mathrm{inter}}_{ij}$, or between levels with the same energy sign, i.e. intraband transitions described by $\sigma^{\mathrm{intra}}_{ij}$.
For temperatures $T\sim 300 K$ and finite chemical potential $\mu \sim 100$ meV, the intraband transitions have to be taken into
account \cite{HipPRB12}.
Indeed, as discussed below, the largest values for the optical conductivity are associated with the intraband transitions.

\subsection{Periodic Deformations along a Single Direction \label{1D-periodic}}

For a deformation along a single direction where $ z = h(x)$, the velocity operators can be read from (\ref{Eq:vel}) and are given
by
\be
v_x=\frac{\alpha_x}{\sqrt{1 + h^2_x}},
\qquad\mbox{ and }\qquad
v_y=\alpha_y  \label{vy} \ .
\ee
The operator $v_x$ acquires the factor $1/\sqrt{1 + h^2_x}$, which can be interpreted as the projection of the unitary vector
$\hat e_x$ into the deformed graphene surface. For a general two-dimensional deformation, such a simple geometric interpretation
is not possible.

The calculation of the real part of the longitudinal optical conductivity $\Re\,\sigma _{ii}$ requires the matrix elements (m.e.'s)
of the velocity operator, $\langle\vec k^\prime, s^\prime |v_i| \vec k,s\rangle$, between eigenstates of the hamiltonian with positive
$(s=+)$ and negative $(s=-)$ eigenvalues; see Eq. (\ref{Sol:geral}) for definition of the spinors.
For a periodic deformation, the quantisation condition (\ref{Eq:quantizacao}) determines
\be
k_x \equiv k_r = \frac{2\pi \, r \, \omega}{ \xi }  \qquad \mbox{ and } \qquad \vec k=k_r\hat e_x+k_y \hat e_y\, .\label{kr}
\ee
Translation invariance along the $y$ direction implies that the m.e.'s are diagonal in $k_y$, but mix different values of $r$ and $r'$.
The energy conservation condition $\Omega_{mn}=\Omega$  in (\ref{kubo1}) gives us
\be
k_y^2=\frac{\Omega^2}{4}-\frac{1}{2}\left(k_{r^\prime}^{2}+k_r^2\right)+
\frac{(k_{r^\prime}^{2}-k_r^2)^2}{4\Omega^2}-M^2 \ . \label{kysol}
\ee
For interband transitions, the solution of the energy conservation constraint
\be
  \sqrt{k_r^2+k_y^2+M^2}+\sqrt{k_{r^\prime}^{ 2}+k_y^2+M^2}=\Omega,  \label{interom}
\ee
requires $\Omega^2\,>\,|k_r^2-k_{r^\prime}^2|$ and its solution is given by (\ref{kysol}).
On the other hand, for intraband transitions, energy conservation reads
\be
\sqrt{k_r^2+k_y^2+M^2}-\sqrt{k_{r^\prime}^{ 2}+k_y^2+M^2}=\Omega ,  \label{intraom}
\ee
and, therefore, $\Omega^2 < k_r^2 - k_{r^\prime}^2$ with $k_y$ given by (\ref{kysol}).

\begin{figure}[t] 
   \centering
   \includegraphics[scale=0.3]{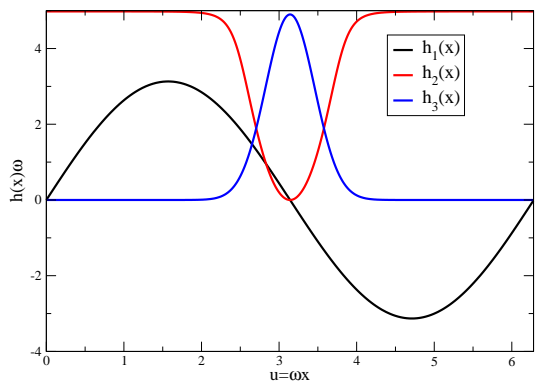}
   \caption{The profiles $h_1(x)$, $h_2(x)$ and $h_3(x)$ defined in (\ref{Eq:various_h}).}
   \label{fig:FuncaoF_profile}
\end{figure}

In the following, we will show results for the following geometric deformations
\begin{equation}
 \begin{array}{lll}
  h_1 (x) = \frac{A}{\omega} \,  \cos u  &A = 3.13 ,  & \\
   & &   \\
  h_2 (x)  =   \frac{A}{\omega} \, \left( \int^u_\pi \frac{dv}{(v - \pi)^8 + b} \right)^2   & A = 0.05, & b = 0.05 , \\
   & &  \\
  h_3 (x)  =  \frac{A}{\omega} \, \exp\left\{ - \frac{ ( u -  \pi )^2}{2b} \right\}  & A = 4.9, & b = 0.1
  \end{array}
  \label{Eq:various_h}
\end{equation}
where $u = \omega \, x$, $2 \pi /\omega$ is the spatial periodicity of the deformation. Note that for the geometries defined
by $h_2(x)$ and $h_3(x)$ the first derivative have a discontinuity at $ x = 0$, $\pm 2 \pi /\omega$,  $\pm 4 \pi /\omega$, $\dots$
but the square of its second derivative, which enters in the definition of $\xi$, is always a continuous function of $x$. The geometric
profiles associated with $h_1(x)$, $h_2(x)$ and $h_3(x)$ can be seen in Fig.~\ref{fig:FuncaoF_profile}.

The examples provide by the profiles $h_1(x)$, $h_2(x)$ and $h_3(x)$ show that optical conductivity is sensitive to the deformation
induced in the graphene surface. Of course, this rises the question of optimising the geometry to either tune the value of the
conductivity or maxima/minimise $\sigma_{ii}$. Although these are important questions from the pratical point of view, they will not be
discussed here. However all the necessary expressions are provided and the interested reader can use them to tune, maximise or
minimise the optical conductivity.

\subsubsection{Interband transitions}

For interband transitions, the matrix elements of the velocity operator depend only on the discrete labels $r$ and $r^\prime$;
see Eqs.~(\ref{kr}) and (\ref{kysol}) for definitions.
Let us introduce the shorthand notation for the m.e.'s
$\langle \vec k^\prime,+|v_x|\vec k,- \rangle\equiv v^{x,\mathrm{inter}}_{r^\prime r}$ and an
analogous form for the m.e.'s of $v_y$. It follows that
\begin{eqnarray}
 v^{x,\mathrm{inter}}_{r^\prime r} & = & -i \, \frac{G^{+-}_{r^\prime r}}{2\xi} \,
                                                           \left(\frac{ik_{r^\prime}-k_y}{E_{k^\prime}+M}-\frac{ik_r+k_y}{E_k-M}\right), \label{vxinter} \\
  v^{y,\mathrm{inter}}_{r^\prime r} & = &  \, \delta_{rr^\prime}  \, \frac{Mk_y+ik_rE_k}{E_k\sqrt{E_k^2-M^2}}, \label{vyinter}
\end{eqnarray}
where
\be
  G_{r^\prime r}^{\lambda^\prime \lambda} =
      F(r-r^\prime) \,\sqrt{\left(1+ \lambda \frac{M}{E_k}\right)\left(1+ \lambda^\prime \frac{M}{E_{k^\prime}}\right) }\, .
\ee
The information on the geometry of the graphene sheet is summarized in the function
\be
F(t) = \int_0^{2\pi}du \, e^{ i \phi(t,u)} ,  \label{f1}
\ee
through the phase-factor
\be
\phi(t,u)=2\pi\frac{t}{\xi}\int^u_0 dv \sqrt{ 1 + h^2_x(v )} \label{f2} \ .
\ee
In the absence of deformation, i.e. for $h = 0$, the function $F(t)$ reduces to $2\pi \delta_{t \, , \, 0}$ as expected from the
translational invariance along $x$, modulo the periodic boundary condition.
The function $F(t)$ encodes the information on the geometry of the graphene sheet and shape the contribution of both
inter- and intra-band transitions to the optical conductivity.

\begin{figure}[t] 
   \centering
   \includegraphics[scale=0.3]{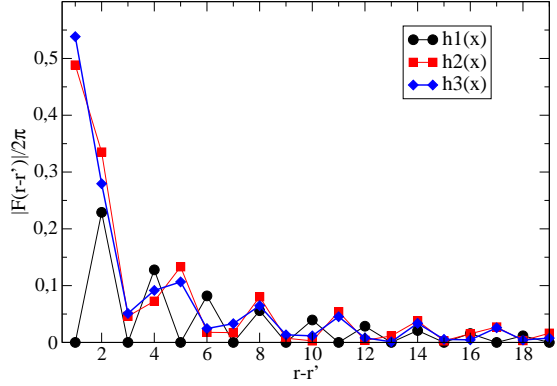}
   \caption{The $| F(r - r^\prime) |/(2\pi)$ function for the profiles defined in (\ref{Eq:various_h}).}
   \label{fig:FuncaoF}
\end{figure}

The function $| F(t) |$ is plotted on Fig.~\ref{fig:FuncaoF} for the periodic geometric deformations given in
(\ref{Eq:various_h}). The profiles on Fig.~\ref{fig:FuncaoF} have the same arc length $\xi / 2 \pi= 2.29$.

The real part of the optical conductivity along the deformation profile is given by:
\be
\frac{\Re \, \sigma_{xx}^{\mathrm{inter}}}{\sigma_0} =
    -\frac{8\omega}{\Omega \xi} \, \sum_{r,r^\prime} \,
\Delta ^{+-}_{r^\prime r} \, D_{r^\prime r}^\Omega  \, |v^{x,\mathrm{inter}}_{r^\prime r}|^2, \label{rsxxinter}
\ee
where $\sigma_0=e^2/4$ is the bulk value,
\be	
D^\Omega_{r^\prime r}=\frac{\left|1-\left(\overline k_{r^\prime}^2-\overline k_r^{2}\right)^2\right|}{\sqrt{1-2 \left(\overline k_{r^\prime}^{2}+\overline k_r^2+2\overline M^2\right)+\left(\overline k_{r^\prime}^{2}-\overline k_r^2\right)^2}}, \label{eq:Drrl}
\ee
with the dimensionless quantities $\overline k_r=k_r/\Omega$, $\overline M=M/\Omega$ and $\xi$ is defined in
(\ref{Eq:quantizacao1}). $k_y$ is given by (\ref{kysol}) and it satisfies the energy conservation condition (\ref{interom}).
The difference between the Fermi-Dirac functions is defined as
\be
\Delta ^{\lambda^\prime \lambda}_{r^\prime r}=f(\lambda^\prime\, E_{k^\prime}-\mu) - f( \lambda \, E_k-\mu) \ .
\ee

The real part of the optical conductivity along the $y$ direction reads
\be
\frac{\Re \,\sigma_{yy}^{\mathrm{inter}}}{\sigma_0} = -\frac{\omega}{\Omega\xi} \, \sum_r  \,
      \frac{\Delta^{+-}_{r\,r} \,
      \left[    \overline M^2\left(\frac{1}{4}-\overline k_r^2\right)+\frac{1}{4}\overline k_r^2\right]}{\sqrt{\frac14-\overline k_r^2-\overline M^2}
      \, \left(\frac14-\overline M^2\right)^2} \, . \label{rsyyinter}
\ee
From its definition it follows that the dimensionless optical conductivity is a function of the dimensionless
quantities $\beta \Omega$, $\omega / \Omega\xi$, $\mu / \Omega$ and $M / \Omega$, i.e.
\be
\frac{\Re \,\sigma_{yy}^{\mathrm{inter}}}{\sigma_0} =  \mathcal{S}_{yy}
\left( \beta \,  \Omega \ , \  \frac{\omega }{ \Omega\,\xi }  \  ,  \  \frac{\mu }{ \Omega} \ , \ \frac{M}{\Omega} \right) \ .
\label{Eq:syy_rescaling}
\ee
This type of behavior is not seen in $\Re \,\sigma_{xx} / \sigma_0 $ which encodes the information on the geometry of the graphene
surface via the function $F(t)$.
The scaling properties of $\Re \,\sigma_{yy} / \sigma_0 $ summarised in Eq. (\ref{Eq:syy_rescaling})
simplifies considerably the analysis of the numerics.

The optical conductivities given by (\ref{rsxxinter}) and (\ref{rsyyinter}) have singularities, the Van Hove singularities,
associated with the zeros of denominator of $D^\Omega_{r^\prime r}$. For zero electron mass, the Van Hove singularities do not
contribute to $\Re \,\sigma_{xx} / \sigma_0$ as they are cancelled by the vanishing of the m.e.'s of the velocity.
This is no longer valid in the case of $M \ne 0$.
The cancellation does not occur in $\Re \,\sigma_{yy} / \sigma_0$ and, therefore, the discrete spectrum of $k_x$,
due to the periodic boundary conditions, gives a nonvanishing contribution to the conductivity.
However, $\Re \,\sigma_{yy} / \sigma_0 \rightarrow 1$ as one approaches the bulk graphene limit, i.e. when
$\omega \rightarrow 0$.

For real graphene, the Van Hove singularities are washed out by disorder and imperfections in the graphene and in the
periodic deformation. In what concerns the optical conductivity, this can be simulated by performing an average in the
parameters of the model, see e.g.~\cite{HipPRB12}.   The results reported below for $\sigma^{\mathrm{inter}}_{yy}$ and $\sigma^{\mathrm{intra}}_{xx}$  are
 an average over $\omega$, assuming a gaussian distribution of frequencies with standard deviation of $\omega/10$.
For $\sigma^{\mathrm{inter}}_{xx}$ such effects are tiny and there is no need to average the results.

\subsubsection{Intraband transitions}

For intraband transitions the m.e.'s of $v_y$ are diagonal in the momentum states $\vec k$ and, therefore,
$\sigma^{\mathrm{intra}}_{yy} = \sigma^{\mathrm{intra}}_{xy} = 0$.

The conductivity $\Re \, \sigma_{xx}^{\mathrm{intra}}$ is given by (\ref{rsxxinter}) after replacing the m.e.'s of $v_x$ by
\be
v^{x,\mathrm{intra}}_{r^\prime r,\lambda} =\frac{iG^{\lambda\lambda}_{r^\prime r}}{2\xi}\left(\frac{-ik_{r^\prime}+k_y}{\lambda E_{k^\prime}+M}+\frac{-ik_r-k_y}{\lambda E_k+M}\right)\ .   \label{vxintra}
\ee
and $\Delta^{+-}_{r^\prime r} \rightarrow \Delta ^{\lambda\lambda}_{r^\prime r}$, with $\lambda=\pm 1$.
It follows that
\be
\frac{\Re \, \sigma_{xx}^{\mathrm{intra}}}{\sigma_0} = -\frac{8\omega}{\Omega \xi} \, \sum_{r,r^\prime,\lambda}
\Delta ^{\lambda\lambda}_{r^\prime r} \, D_{r^\prime r}^\Omega \, |v^{x,\mathrm{intra}}_{r^\prime r,\lambda}|^2\, .\label{rsxxintra}
\ee

For flat graphene $r^\prime=r$ and the Pauli Principle forbids a transition to the same state, therefore,
the intraband optical conductivity vanishes. On the other hand, if the graphene sheet is deformed, $r^\prime \ne r$ transitions
are allowed and $\Re \, \sigma_{xx}^{\mathrm{intra}} \ne 0$.

In what concerns the dependence with the chemical potential, an increase in $\mu$ means a larger phase space for
intraband transitions and, therefore, an increase in the intraband optical conductivity. In addition,
given that the number of possible electronic interband transitions is determined mainly by the chemical potential, $\mu$ also determines
the contribution of the Van Hove singularities to $\Re \,\sigma_{xx} / \sigma_0$.

From its definition, it follows that
\be
\frac{\Re \,\sigma_{xx}}{\sigma_0} =  \mathcal{S}_{xx}
\left( \beta \, \omega \ , \  \frac{\Omega }{ \omega }  \  ,  \  \frac{\mu }{ \omega} \ , \ \frac{M}{\omega}  \ , \  \omega \, h(\omega x)  \right) \ .
\label{Eq:sxx_rescaling}
\ee
This relation applies both to the contribution of the intra and interband transistions to the optical conductivity.

\subsection{Numerical Results \label{results}}

In this section, we report on the numerical results for the optical conductivity for inter, see Eqs. (\ref{rsxxinter}) and (\ref{rsxxintra}),
and intraband transitions, see Eq. (\ref{rsyyinter}), considering the geometric profiles defined in (\ref{Eq:various_h}).
In what follows, the Dirac electron is supposed to be massless, the spatial periodicity $2\pi/\omega$ is varied between 30 and 200 nm
and the chemical potential is varied in the range 40 - 450 meV for a temperature $T=300$ K.

As described below, we have observed quite different behaviors for the conductivities associated with the inter and intraband
transitions. The intraband transitions give much larger contributions to the optical conductivity in the far infrared region, where
it dominates for $\Omega \sim \omega$ and for a finite chemical potential. On the other hand, as $\Omega$ increases
the contributions to the conductivity approach its asymptotic values and the larger values are now associated with the
interband transitions.

\begin{figure}[t] 
   \centering
   \includegraphics[scale=0.30]{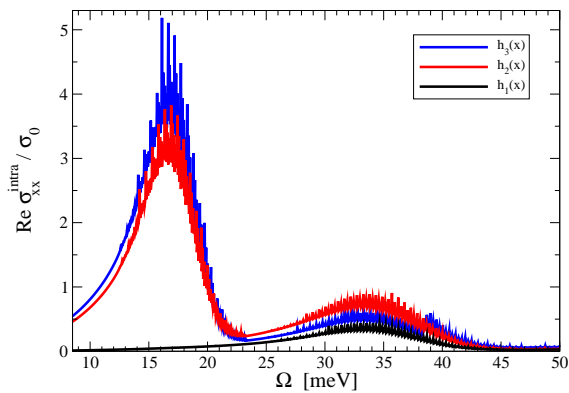}
   \caption{{\it Color on-line.} Optical conductivity for intraband transitions for the geometries (\ref{Eq:various_h}), a
                chemical potential $\mu=250$ meV, $2\pi/\omega=100$ nm and the same arc length $\xi$.}
   \label{figcompintral}
\end{figure}

Let us start looking to the contribution of the intraband transitions to the optical conductivity.
Fig. \ref{figcompintral}  shows the intraband optical conductivity for the geometries defined in (\ref{Eq:various_h})
computed for a chemical potential $\mu=250$ meV, a spatial periodicity $2\pi/\omega=100$ nm and the same arc length $\xi$.
The behavior of
$\Re \sigma_{xx}^{\mathrm{intra}}$ is determined mainly by the Van Hove singularities and is modelled by the geometry of the
graphene sheet. In the limit of $\omega\rightarrow\infty$ and $\xi\rightarrow0$ the first peak of Fig. \ref{figcompintral} narrows and move towards $\Omega=0$, becoming the Drude peak, that's is expected for a infinite and disorder-free system \cite{HipPRB12}.

The Van Hove singularities in $\Re\sigma_{xx}$ are defined by the zeros of the denominator in Eq. (\ref{eq:Drrl})
\begin{displaymath}
  D^\Omega_{r^\prime r} = \frac{ ~ \left|1-\left(\overline k_{r^\prime}^2-\overline k_r^{2}\right)^2\right|  ~}{ \overline k_y } \ ,
\end{displaymath}
 which occur for $|\overline k_r|-|\overline k_{r^\prime}|=1$, see Eq. (\ref{intraom}),  and $r \ne r^\prime$.
Given that the m.e.'s of $v_x$ and the denominator of $D^\Omega_{r^\prime r}$ do not vanish simultanously,
the positions of the Van Hove peaks happen for $\Omega=2\pi\omega \, s \, / \, \xi$, where
$ s  = | r -  r^\prime| = 1$, $2$, $3$ $\dots$

Fig. \ref{figcompintral} shows a peak structure at $\Omega = 17 \, s$ meV.
The peaks are associated with the Van Hove singularities and each Van Hove peak is modulated by the corresponding
$|F(\pm s)|^2$ function.  As can be seen in Fig. \ref{fig:FuncaoF}, $|F(\pm s)|^2$ decreases fast with $s$ and, therefore,
the higher peak intensity occurs for smaller $s$. Indeed, the first peak corresponds to $s=1$, and the successive ones to
$s = 2$, $3$, $\dots$.
For the profile function $h_1(x) = A \cos \omega x$, the function $F(\pm s)$ vanishes at odd $s$ and, for
this particular geometry, there are no peaks for odd $s$.

The maxima of the various peaks are proportional to $|F(\pm s)|^2$ and, therefore, its relative
value is a signature of the function $|F(\pm s)|^2$. This can also be read comparing Figs. \ref{fig:FuncaoF} and \ref{figcompintral}.

The behavior of $F(\pm s)$ at large $s$ also explains why the contribution of the intraband transitions to the optical
conductivity vanishes in the limit of $\Omega \gg \omega$. Indeed, for sufficiently large $\Omega$ or small $\omega$,
Eq. (\ref{intraom}) implies that $k_r \gg k_{r^\prime}$ and, therefore, $F(r  - r^\prime)$ is essentially zero as $r  - r^\prime$
becomes large.

\begin{figure}[t] 
   \centering
  \includegraphics[scale=0.30]{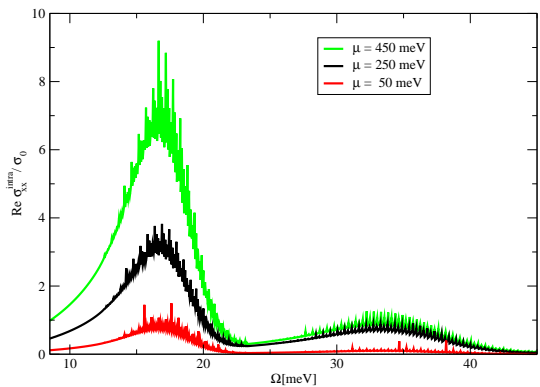}
   \caption{{\it Color on-line.} Optical conductivity associated with intraband transitions for the Gaussian geometry defined
                 in (\ref{Eq:various_h}) with a $2 \pi/\omega=100$ nm and for different chemical potentials $\mu$.
                 Note that the widths become larger as $\mu=$ increases from 50 meV to 450 meV.}
   \label{figmugauss}
\end{figure}

\begin{figure}[t] 
   \centering
  \includegraphics[scale=0.35]{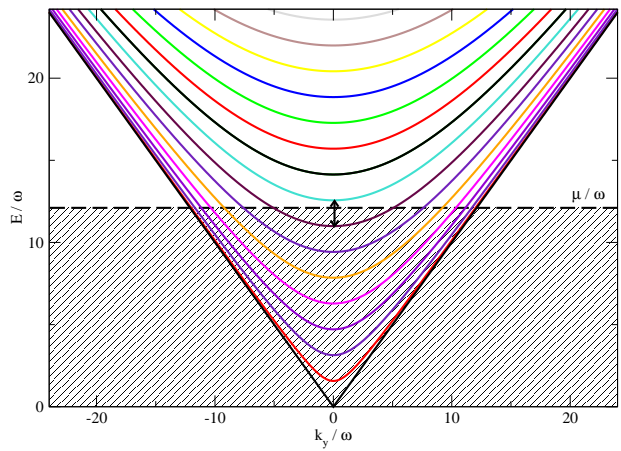}
   \caption{{\it Color on-line.} Band structure for the Gaussian profile defined in (\ref{Eq:various_h}). For a given chemical potential $\mu$
                 (dashed line), the possible intraband transitions can only occur from the shaded area to states above the dashed line.
                 The dominant contribution to $\Re \sigma_{xx}^{\mathrm{intra}}$ is represented by the arrow. }
   \label{figenergias}
\end{figure}

The widths of various peaks in the optical conductivity associated with the intraband transitions depend on the chemical potential $\mu$.
For example, for the Gaussian profile, see (\ref{Eq:various_h}), the quantisation condition (\ref{Eq:quantizacao}) provides the
band structure of Fig. \ref{figenergias}. For a given chemical potential $\mu$ (dashed line), the intraband transitions that can contribute
to the optical conductivity can only occur from and to the shaded area.
The dominant contribution to $\Re \sigma_{xx}^{\mathrm{intra}}$ is represented by the arrow in Fig. \ref{figenergias};
see~\cite{HipPRB12} for a discussion.

The phase space for the intraband transitions increases with $\mu$ and, therefore, the corresponding widths also increase with
the chemical potential.

Fig. \ref{figmugauss} illustrates the effect  of changing the chemical potential for the Gaussian profile with $2\pi/\omega=100$ nm.
As shown, the positions of the peaks are found at $\Omega = s \, 17\,$meV and their width increases with  $\mu$.
Similar curves can be drawn for the other profiles in (\ref{Eq:various_h}).

\begin{figure}[t] 
   \centering
   \includegraphics[scale=0.35]{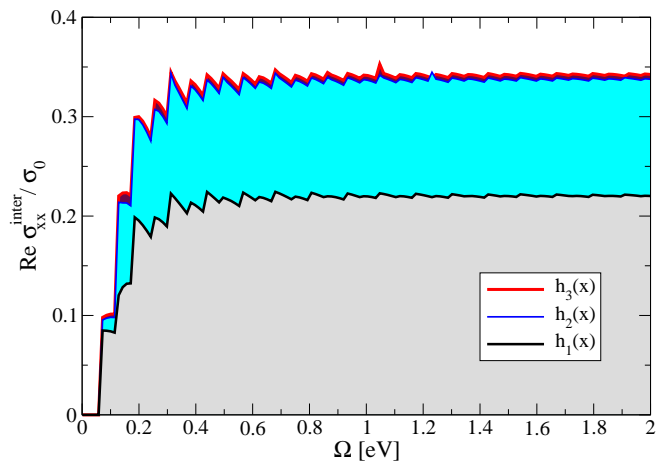}
   \caption{{\it Color on-line.} Contribution of the interband transitions to the optical conductivity for different geometries with  $\mu=40$ meV and a fixed periodicity of $2\pi/\omega=100$ nm. From bottom to top one sees $\Re \sigma_{xx}^{\mathrm{inter}}/\sigma_0$ computed with $h_1(x)$ (gray), $h_2(x)$ (blue), and $h_3(x)$ (red)  defined in (\ref{Eq:various_h}). }
   \label{figsigxxinter}
\end{figure}

\begin{figure}[t] 
   \vspace{0.5cm}
   \centering
   \includegraphics[scale=0.35]{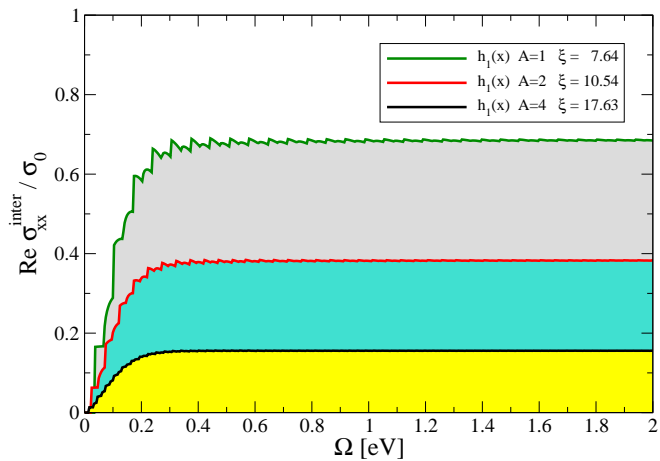}
   \caption{{\it Color on-line.} Contribution of the interband transitions to the optical conductivity for the geometric profile
                 $A\cos \omega x$ for different amplitudes.  From top to bottom $\Re \sigma_{xx}^{\mathrm{inter}}/\sigma_0$ follows the
                 sequence  $A\,\omega=$ (1, 2, 4)  and $\xi=$(7.64, 10.54, 17.63).
                 The periodicity is given by $2\pi/\omega= 100$ nm and the chemical potential is fixed at 40 meV. }
   \label{figsigxxintercos}
\end{figure}

\begin{figure}[t] 
\vspace{0.5cm}
   \centering
   \includegraphics[scale=0.35]{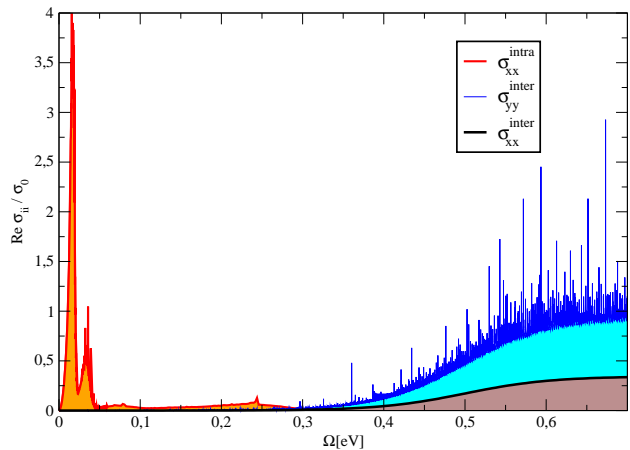}
   \caption{{\it Color on-line.} Optical conductivity for the Gaussian geometric profile given in (\ref{Eq:various_h})
                  with $2\pi/\omega=100$ nm and $\mu=$ 250 meV. The low frequency peak in the far infra-red comes from
                  $\Re \sigma^{\mathrm{intra}}_{yy}$ (green).
                  Each contribution of interband transitions to the optical conductivity presents a plateaux around $\sigma_0$ for
                  $\Re \sigma^{\mathrm{inter}}_{yy}$ (brown) and about $0.3\sigma_0$ for $\Re \sigma^{\mathrm{inter}}_{xx}$ (blue). }
   \label{figtodos}
\end{figure}

Let us turn our attention to the optical conductivity associated with interband transitions. Recall that the matrix elements of $v_x$
vanish for the interband transitions at the pole frequency, see Eq. (\ref{eq:Drrl}),
and, in this way, the Van Hove singularities do not contribute to $\Re \sigma_{xx}^{\mathrm{inter}}$.
Indeed, as can be seen in Figs.~\ref{figsigxxinter}, \ref{figsigxxintercos}
and \ref{figtodos}, $\Re \sigma_{xx}^{\mathrm{inter}}$ has no peak structure.

The maxima of the peaks associated with the Van Hove singularities, i.e. the pole contribution to the optical conductivity in
$\Re \sigma_{xx}^{\mathrm{intra}}$ look rather different than the corresponding contribution for $\Re \sigma_{yy}^{\mathrm{inter}}$.
If for the former quantity, see Eq. (\ref{rsxxintra}), there is a double sum in $r$ and $r^\prime$ and the frequency $\Omega$
contributes to the conductivity whenever $\Omega =  2 \pi \, s \, \omega / \xi$ for $s = |r  - r^\prime|$,
for the later one, see Eq. (\ref{rsyyinter}), it involves a single sum over $r$ and only the frequency
$\Omega = 4 \pi \, | r | \, \omega / \xi$ contributes to the conductivity. Then, for a given $\Omega$ associated with a pole on
the optical conductivity, $\Re \sigma_{xx}^{\mathrm{intra}}$ adds the contribution of an infinite number of poles, while
$\Re \sigma_{yy}^{\mathrm{inter}}$ sees only a single pole and one expects higher peak values for $\Re \sigma_{xx}^{\mathrm{intra}}$
when compared with $\Re \sigma_{yy}^{\mathrm{inter}}$. Note that the number of frequencies associated with the Van Hove singularities
is determined by the chemical potential.
Furthermore, the widths of the peaks associated with interband and intraband transitions is expected to be distinct, with
Van Hove peak intraband transitions showing a larger width. The diverse widths of the Van Hove peaks for the intra and interband
transitions was already discussed in \cite{HipPRB12} for graphene nanoribbons.

The optical conductivity $\Re \sigma_{xx}^{\mathrm{inter}}$ associated with the interband transitions for different geometries,
a chemical potential $\mu=40$ meV, a fixed spatial periodicity $2\pi/\omega=100$ nm and for profiles having the same
arc length $\xi$ is shown in Fig. \ref{figsigxxinter}.
As expected no Van Hove singularities are observed and the conductivity approaches a plateaux whose value depends on the
particular geometry. Note, however, that the values for $\Re \sigma_{xx}^{\mathrm{inter}}$ are not that sensitive to the
graphene profiles and are significantly smaller than the corresponding optical conductitvity for the intraband transitions -- see
Figs.~\ref{figmugauss} and \ref{figcompintral}.

The effect on the interband conductivity of changing the arc length $\xi$ for the profile $h(x) = A\cos\omega x$ can be seen
in Fig. \ref{figsigxxintercos}.  The curves are for $2\pi/\omega= 100$ nm, a chemical potential of 40 meV and
$(A\,\omega , \xi ) = (1 , 7.64)$, $(2 , 10.54)$, $(4 , 17.63)$. As Fig. \ref{figsigxxintercos} shows, a larger $\xi$ means a smaller
$\Re \sigma_{xx}^{\mathrm{inter}}$. This behavior with $\xi$ can be understood as the matrix elements in
$\Re \sigma_{xx}^{\mathrm{inter}}$  scale as $1/\xi^2$.

In Fig. \ref{figtodos} the contribution of the various transitions to the optical conductivity are illustrated for the case of a
Gaussian profile with spatial periodicity $2\pi/\omega=100$ nm and chemical potential $\mu=$ 250 meV.
The highest conductivity occurs for $\Re \sigma^{\mathrm{intra}}_{xx}$, detailed in Fig. \ref{figmugauss}, and occurs for low frequency,
i.e. in the far infrared region of the spectra.

The conductivity associated with interband transitions show a plateaux as $\Omega$
increases. As $\Omega \to \infty$,  the effects due $T$, $\mu$ and the periodic boundary condition can be disregarded and, as
clearly seen in the figure, the bulk value is approached for ${\Re \,\sigma_{yy}^{\mathrm{inter}}}/{\sigma_0} \rightarrow 1$.
In what concerns the high frequency limit for $\Re \sigma^{\mathrm{inter}}_{xx}$, its plateaux value correlates with the geometry of
the graphene sheet and it can be shown that
\be
   \frac{\Re \sigma^{\mathrm{inter}}_{xx}}{\sigma_0} \longrightarrow
    \left( \frac{2 \pi}{\xi} \right)^2 +  \frac{2}{\xi^2} \sum^{\infty}_{t = 1} \left| F(t) \right|^2
\ee
as $\Omega \to \infty$.

\section{Summary \label{ultima_sec}}

The present work make use of the geometric framework to investigate electrons in graphene.
In particular, we look at in-plane and out-of-plane displacements as deformations and the corresponding Dirac equation
in curved 2D space is derived. We focus on static metric that ensures an hermitian Hamiltonian \cite{Huang2009}.

A general deformation along a single spatial direction is investigated and we are able to provide an analytical solution of the
corresponding Dirac equation in a curved space.
The solution shows that, for this class of geometries, curvature induces an extra phase in the electron wave
function. This opens the possibility of engineering Aharonov-Bohm interference type devices with geometric deformations of the
graphene sheet in one direction.

An experimental realisation of deformations along one direction was already achieved bending
graphene nanoribbons \cite{Koch2012}. This allows to hope that, in the future, one can tailor the geometry of graphene wafers
to create controlled devices with tuned conductance properties. Along this lines, we have considered an idealised apparatus
to explore Aharonov-Bohm interferences based on geometrical deformations in one spatial direction.

For periodic deformations a quantisation rule for the energy is found from the phase structure. This suggests the presence of
electron Block waves in the background of a ``crystal structure" created by the periodicity of the deformation, which is
illustrated for a periodic out of plane ripple.

The Sturm-Liouville type of equation associated with the Dirac equation in curved space was obtained for the ripple, illustrating
the possibility of mapping complex potentials into solvable geometric problems.

The induced deformations  of the graphene/nanorribon surface has the potential to modify the charge carriers properties,
namely the associated electronic states and optical properties.
For periodic out-of-plane deformations, the quantisation rule for the energy eigenvalues, the phase structure of the eigenstates of the
Dirac Hamiltonian on the surface of the curved graphene, change in a sensible form the optical conductivity.
Different examples of periodic deformations along the arbitrary $x$-direction shows an enhancement of the optical conductivity in
the far and mid infrared frequency range for periodicities $\sim100\,$nm. The enhancement of the optical conductivity comes
mainly from the intraband transitions and is associated with the Van Hove singularities modulated by the geometry of the
graphene sheet. The difference of the contributions of the Van Hove singularities to the interband and intraband optical conductivities
explains the observed scales for $\sigma_{ii}$. Furthermore, we have shown that the manipulation of the profile functions can alter both
the width and position of the peaks. This opens the problem of finding the geometric profile to maximise the optical conductivity or
to tune the geometric profile and produce a given value for the conductivity. We have not explored these questions here, but all
the expressions derived allow to investigate such important problems.

As reported on Sec.~\ref{results}, the optical conductivity of a curved graphene sheet depends on the polarization of the photon
relative to the direction along which the graphene is deformed. This anisotropy in the optical conductivity can be explored to
create a dichroism effect.
For a linearly polarized incident photon with momenta perpendicular to the graphene sheet and polarization angle
$\theta$, measured relative to the $x$-axis, the rotation of the polarization angle in the graphene sheet is given by
\be
\tan \theta^\prime=\frac{2+Z \sigma_{xx}}{2+Z \sigma_{yy}}\tan \theta
\label{Eq:Impedancia}
\ee
where
$Z$ is the medium impedance and $\theta^\prime$ is the polarization angle of the refractive electromagnetic wave.
In the above expression we have neglected the contributions from $\sigma_{xy}$.
For a graphene sheet in vacuum, $Z_0 \, \sigma_0 \approx 0.013 \ll 1$, where $Z_0 = \sqrt{\mu_0/\varepsilon_0} \approx 377 \, \Omega$
is the vacuum impedance, and $\theta^\prime \approx \theta$. Further, if $\sigma_{xx} = \sigma_{yy}$ the polarization angle remains
essentially unchanged and $\theta^\prime = \theta$.
For photon frequencies of the order of $\omega$, the large values of $\Re \sigma_{xx}^{\mathrm{intra}}$ relative to
$\Re \sigma_{yy}^{\mathrm{inter}}$, makes graphene opaque for photons polarized along the $x$-axis, i.e. graphene is
a linear polarizer for frequencies $\Omega \approx \omega$. For such frequencies the dichroism is rooted on (i)
the Van Hove singularities arising from the periodicity of the deformation,
(ii) the break of the translational invariance along the direction of the deformation ($x$) and subsequent non-diagonal optical transitions
and
(iii) the chemical potential and temperature to (de)populate states of the (valence)conduction band.
As the photon frequency $\Omega$ increases, promptly $\Re \sigma_{xx}^{\mathrm{intra}}$ becomes negligible compared to the
contribution to the optical conductivity associated with the interband transitions.
For photon frequencies $\Omega \gtrsim \mu$, the optical conductivity is dominated by the interband transitions
and $\Re \sigma_{xx}^{\mathrm{inter}}$ is suppressed, relative to $\Re \sigma_{yy}^{\mathrm{inter}}$, by the factor $(\xi/2\pi)^2$.
For such frequencies, the optical response of graphene to linear polarized light can be changed modifying only the arc length $\xi$
and, in this way, a dichroism effect is produced. Furthermore, optical dichroism can be enhanced putting the graphene sheet on
medium with a large impedance, see Eq. (\ref{Eq:Impedancia}),  like e.g. a magnetic medium, with $\mu\gg\mu_0$, or
placing it inside a waveguide -- see \cite{HipPRB12} for discussions.

\section*{Acknowledgements}

The authors acknowledge financial support from the Brazilian
agencies FAPESP (Funda\c c\~ao de Amparo \`a Pesquisa do Estado de
S\~ao Paulo) and CNPq (Conselho Nacional de Desenvolvimento
Cient\'ifico e Tecnol\'ogico). O. Oliveira acknowledges financial support from FCT under contracts
PTDC/\-FIS/\-100968/\-2008 under the initiative QREN financed by the UE/FEDER through the Programme
COMPETE.

\begin{appendix}

\section{Out-of-Plane Displacements: the general case \label{Ap:out}}

In order to fix our notation and conventions, here we give the details on the Dirac equation on a curved 2D space.
An analogous formalism was also presented in \cite{Kerner2012}.

Let us consider a two-dimensional out-of-plane displacement
parametrized as
\begin{displaymath}
 z = h(x,y) \ .
\end{displaymath}
The associated metric reads
\be
\left( g_{\mu\nu} \right) = \left(
\begin{array}{ccc}
  1 & 0 & 0 \\
 0 & -h_x^2 -1              &   -h_x  \, h_y\\
 0 &  -h_x  \, h_y               & -h_y^2-1
\end{array}
\right)
\ee
where $h_x$ and $h_y$ stands for the partial derivative with respect to $x$ and $y$.
Defining the local Lorentzian frame by
$ds^2=\eta_{AB}\th^A\th^B$, where $\th^A = e^{A}_{\mu}dx^{\mu}$, one can write
\bea
\th^0&=&dt \ , \nonumber\\
\th^1&=&A \, dx + B  \, dy \ ,\nonumber\\
\th^2&=&C \, dx + D \, dy \ .
\eea
Rewriting $ds^2$ in terms of $A$, $B$, $C$ and $D$ we get
\bea
A^2 +C^2        & = & 1+h_x^2 \ ,\nonumber\\
B^2 +D^2        & = & 1+h_y^2 \ ,\nonumber\\
A \, B + C \, D & = & h_x h_y \, .
\eea
This system have many solutions but we will consider those which are $x$ and $y$ symmetric and assume
$B=C$. This is not sufficient to define a unique solution. Indeed, the system has four different possible solutions
but, in the following, we will take
\bea
 A       &=&    \frac{h_y^2 + h_x^2 \sqrt{1 + h_x^2 + h_y^2} }{h_x^2 + h_y^2} \ , \nonumber\\
 B       &=&    \frac{- h_x h_y + h_x h_y\sqrt{1 + h_x^2 + h_y^2} }{h_x^2 + h_y^2} \ ,\nonumber\\
 D       &=&    \frac{h_x^2 + h_y^2 \sqrt{1 + h_x^2 + h_y^2}}{h_x^2 + h_y^2} \ .
\eea
The nonvanishing matrix elements of the ``vielbein"
\be
\left( e^{A}_{\mu} \right) = \left(
\begin{array}{lll}
 e^{0}_{0} & e^{0}_{1} & e^{0}_{2} \\
 e^{1}_{0} & e^{1}_{1} & e^{1}_{2} \\
 e^{2}_{0} & e^{2}_{1} & e^{2}_{2}
\end{array}
\right) \ ,
\ee
are
\bea
e^0_0 & = &1 \ ,\nonumber\\
e^1_1 & = & \frac{h_y^2 + h_x^2 \sqrt{1 + h_x^2 + h_y^2} }{h_x^2 + h_y^2} \ , \nonumber\\
e^1_2 & = & \frac{- h_x h_y + h_x h_y\sqrt{1 + h_x^2 + h_y^2} }{h_x^2 + h_y^2} \ ,\nonumber\\
e^2_1 & = & e^{1}_{2}   \ , \nonumber\\
e^2_2 & = & \frac{h_x^2 + h_y^2 \sqrt{1 + h_x^2 + h_y^2}}{h_x^2 + h_y^2} \ .
\eea

The nonvanishing spinconnection components are
\be
 \omega^{ij}_k = - \omega^{ji}_k = \left(- h_i h_{jk} + h_j h_{ki} \right) \tilde\Omega (x,y)
\ee
where
\be
 \tilde\Omega (x,y) =
 \frac{- 1 + \sqrt{1 + h_x^2 + h_y^2} }{\left(h_x^2 + h_y^2\right)\sqrt{1 + h_x^2 h_y^2}} \ .
\ee

Finally, the Dirac equation reads
\bea
&&\left\{  i \gamma^{0}\partial_{0} + i \Gamma^{1}\left(\partial_x+\frac{1}{2}\omega_{1}^{12}\gamma_1\gamma_2\right)\right. \nonumber\\
&&\left. + i \Gamma^{2}\left(\partial_y+\frac{1}{2}\omega_{2}^{12}\gamma_1\gamma_2\right) -m\right\}\Psi=0,
\label{Eq:Hgeral}
\eea
with
\bea
\Gamma^1 &=&
   \frac{1}{h_x^2 + h_y^2}
   \left\{ \left( h_y^2 + \frac{h_x^2}{\sqrt{1 + h_x^2 + h_y^2}}\right) \gamma^1 \right. \nonumber\\
&& \left. - h_x h_y \left(1 - \frac{1}{\sqrt{1+ h_x^2 + h_y^2}}\right) \gamma^2\right\} \label{Gammai}
\eea
and $\Gamma^2$ being obtained from $\Gamma^1$ after interchanging $x \leftrightarrow y$ and $\gamma^1\leftrightarrow \gamma^2$.
The Dirac
equation (\ref{Eq:Hgeral}) can be rewritten in the Hamiltonian form
\be
 i \partial_0 \psi = H \psi
\ee
where
\bea
 H  &=& - i \beta \, \Gamma^{1}\left(\partial_x+\frac{1}{2}\omega_{1}^{12}\gamma_1\gamma_2\right) \nonumber\\
&&       -i \beta \, \Gamma^{2}\left(\partial_y+\frac{1}{2}\omega_{2}^{12}\gamma_1\gamma_2\right)
           + \beta \, m
 \label{Eq:H_Dirac_geral}
\eea
and $\beta = \gamma^0$.


\section{Sturm-Liouville equation: Small and Large Curvature Limits \label{STLE-Mathieu}}

Let us discuss the small and large curvature limits where the Sturm-Liouville equation reduces to a well know differential equation.

In the small curvature limit, i.e. for
$\eta = A \omega \ll 1$,  the second order differential equation (\ref{Eq:SL}) reduces to
the Mathieu equation\cite{Abramowitz}
\be
 \frac{d^2 \phi}{d v^2} + \Big( a - 2 q \cos (2 v ) \Big) \phi = 0
 \label{Eq:MathieuSmall}
\ee
where
\be
  a  =  \zeta \left( 1 + \frac{\eta^2}{2} \right)
  \quad\mbox{and}\quad
  q = \frac{\eta^2}{4} \left( \zeta - 1 \right) \ .
\ee
Note that with such definitions $q$ can take, in principle, any value.
Using a different approach, the limit of the small curvature for a corrugated graphene sheet was also
considered in \cite{Atanasov2010}, where the authors reduced the problem to the solution of a Mathieu
equation \cite{comentario}. The small deformation limit was studied using a geometrical language in
\cite{Kerner2012}, where the authors built the corresponding Dirac equation but did not consider
its general solution, as provided in our work.

The solutions of the Mathieu equation can be classified according to their parity. Using the notation of \cite{Abramowitz},  the parity even corresponds to
$ce_r(v,q)$ for $r = 1, 2, \dots$ and with $a = a_r(q)$, while the parity odd
are associated with $se_r(v,q)$ for $r = 1, 2, \dots$ and with $a = b_r(q)$.
The relations $a = a_r(q)$ and $a = b_r(q)$ define the relation $\zeta = \zeta ( \eta )$ which can be
rewritten in terms of the energy. The solutions $ce_r(v,q)$ and $se_r(v,q)$ have period $\pi$ for
$r$ even and $2 \pi$ for $r$ odd.

For small $q$, it follows that the even parity solutions have a $\zeta$ spectrum given by
\bea
  \zeta^{(+)}_0  & =  & 0 + \mathcal{O} \left( \eta^4 \right) , \nonumber \\
  \zeta^{(+)}_r   & =  & r^2 \left( 1 - \frac{\eta^2}{2} \right)+ \mathcal{O} \left( \eta^4 \right) \quad \mbox{ for } \quad
          r \ge 1 \, .
  \label{Sp_small_plus}
\eea
On the other hand, the $\zeta$ spectrum for the odd parity solutions reads
\be
  \zeta^{(-)}_r  =  r^2 \left( 1 - \frac{\eta^2}{2} \right) + \mathcal{O} \left( \eta^4 \right) \mbox{ for } r \ge 1
  ~ .
  \label{Sp_small_minus}
\ee
Note that the spectra of the periodic solutions of the Mathieu equation reproduce exactly the results found in
(\ref{Eq:quantizacao_small}), where $K_x/ \omega$ takes the role of $\sqrt{\zeta}$.

For large deformations given by $\eta = A \omega \gg 1$, equation (\ref{Eq:SL}) reduces to
a different Mathieu equation
\be
   \frac{d^2 \phi}{d v^2} + \Big( \overline a - 2 \, \overline q \cos (2 v ) \Big) \phi = 0
   \label{Eq:MathieuLarge}
\ee
where
\be
\overline a = 2 \, \overline q = \frac{\zeta \,  \eta^2}{2} \ .
\ee
A similar analysis as for the small curvature limit but relying on the asymptotic series for $a(\overline q)$ give
\be
  \zeta \sim 1/\eta^2
  \label{Eq:zetaLarge}
\ee
for $r \ne 0$ with $\zeta$ vanishing if $r = 0$. As can be seen in Fig. \ref{Fig:quantizacao}, replacing $K_x / \omega$ by
$\sqrt{\zeta}$, the large $\eta$ behaviour is compatible with the prediction of (\ref{Eq:zetaLarge}).

\end{appendix}


\end{document}